\documentclass[onecolumn,showpacs,preprintnumbers]{revtex4}
\usepackage{amsthm, amscd, amsfonts, amssymb, graphicx}
\usepackage{dcolumn}
\usepackage{bm}
\usepackage[english,russian]{babel}

\begin{document}

\title{Model of a hypothetical high-temperature superconductor.
}
\author{K.V. Grigorishin}
\email{gkonst@ukr.net} \affiliation{Boholyubov Institute for
Theoretical Physics of the National Academy of Sciences of
Ukraine, 14-b Metrolohichna str. Kiev-03680, Ukraine.}
\date{\today}

\begin{abstract}
We propose a model of a hypothetical superconductor which includes
impurities with retarded interaction with quasiparticle and
superconducting matrix where these impurity can be applied with
large concentration. Interaction between the impurity and a
conduction electron in such a system has been calculated. We have
found the critical temperature of the system matrix+impurity
essentially exceeds critical temperature of the pure
superconducting matrix.
\end{abstract}


 \maketitle

\section{Introduction}\label{intr}

As it is well known at embedding of impurities in a superconductor
a gap $\Delta$ and an energetic parameter $\varepsilon$ are
renormalized \citep{sad,sad1}. If the superconductor is s-wave
type and impurities are nonmagnetic then the gap and the energetic
parameter $\varepsilon$ are renormalized equally:
$\frac{\widetilde{\Delta}}{\widetilde{\varepsilon}}=\frac{\Delta}{\varepsilon}$,
where $\widetilde{\Delta},\widetilde{\varepsilon}$ are
renormalized values by an impurity scattering. As a consequence a
critical temperature of a superconductor does not change. This
statement is Anderson's theorem. If d-wave pairing takes place or
impurities are magnetic or electrons are paired with nonretarded
interaction (as in BCS theory \citep{fay}) then the gap and the
energetic parameter are renormalized in different ways:
$\frac{\widetilde{\Delta}}{\widetilde{\varepsilon}}<\frac{\Delta}{\varepsilon}$.
Hence the critical temperature decreases and an effect of gapless
superconductivity can take place. In a work \citep{grig1}
generalization of a disordered metal's theory has been proposed
when scattering of quasiparticles by nonmagnetic impurities is
caused with a retarded interaction. The retarded interaction
occurs because the impurities have an internal structure and make
transitions between their states under the action of metal's
quasiparticles. It was shown that in this case embedding of the
impurities in $s$-wave superconductor increases its critical
temperature. The increase of the critical temperature is a
mathematical consequence of an inequality
$\frac{\widetilde{\Delta}}{\widetilde{\varepsilon}}>\frac{\Delta}{\varepsilon}$,
that is the gap and the energetic parameter are renormalized in
the opposite way to a case of magnetic impurities. It should be
notice that the disorder can influence upon phonon and electron
specter in materials. Experiments in superconductive metal showed
suppression of $T_{\texttt{C}}$ by a sufficiently strong disorder
\citep{fiory,hert,bishop,nish}. The strong disorder means that a
free length $l$ is such that $\frac{1}{k_{F}l}\approx 1$ or
$\varepsilon_{F}\tau\approx 1$, where $\tau=l/v_{F}$ is a mean
free time. Collapse of superconducting state takes place near
Anderson's transition metal-insulator, that is when
$\frac{1}{k_{F}l}\gtrsim 1$. In a work \citep{grig2} a
perturbation theory and a diagram technique have been developed
for a disordered metal if interaction of quasiparticles with
impurities is retarded and impurity's oscillations are local.
Transition amplitudes of the impurities between their states under
the action of metal's quasiparticles (an electron-impurity
coupling) have been obtained in the adiabatic approximation.
Eliashberg equations at a critical temperature
$T_{\texttt{C}}^{\ast}$ have been generalized for a case of s-wave
superconductor containing impurities of a considered type. It
found a critical temperature of a system metal+impuryty is more
than a critical temperature of the pure metal
$T^{\ast}_{\texttt{C}}>T_{\texttt{C}}$. In the works
\citep{grig1,grig2} it has been shown that the equations for the
transition temperature $T_{\texttt{C}}^{\ast}$ of the system
metal+impurity can be simplified if we calculate so-called
singularity temperature $T^{\ast}$ only. The singularity
temperature is a superconducting transition temperature if we turn
off the pairing interaction caused by metal's phonon, therefore we
have always $T^{\ast}<T^{\ast}_{\texttt{C}}$. The singularity
temperature can be used as a lower estimate of the critical
temperature of the dirty metal if $T^{\ast}\gg T_{\texttt{C}}$.

In the articles \citep{grig1,grig2} principal possibility of
increasing of the superconductive transitional temperature has
been shown and method of calculation of the transition temperature
was given. However in these articles a model of an impurity was
not proposed. Hence this paper is aimed to propose the model of
the impurity with retarded interaction with quasiparticle, to
propose superconducting matrix where these impurities can be
applied, to calculate the singularity temperature and critical
temperature for the system matrix+impurity.

\section{Model of an impurity.}\label{impur}
\subsection {Structure of the impurity.}\label{impur1}

The impurity must have internal structure that is to have
possibility to be in states with different energies. The simplest
example of this system is a harmonic oscillator. Moreover metal's
quasiparticles must interact with the impurities. We offer the
following construction of the impurity. Large variety of
endohedral complexes - atoms and ions inside the $\texttt{C}_{60}$
cage is known now \cite{elets}. For our aim endoendral complexes
$\texttt{X@C}_{60}$, where $\texttt{X}$ is a noble gas atom
($\texttt{He}$, $\texttt{Ne}$, $\texttt{Ar}$, $\texttt{Kr}$,
$\texttt{Xe}$) trapped in a carbon cage
\cite{breton,cios,saund,rubin,dennis}, are suitable. Since the
noble gas atom has closed electron shell then it does not transfer
charge to the cage and it is in central of the cage (thereby no
breaking the symmetry of the fullerene) unlike metal endoendral
complexes where an atom of metal is shifted to internal surface of
the cage. Moreover the central noble gas atom does not change
chemical properties of a fullerene unlike atom of metal.

\begin{figure}[h]
\includegraphics[width=8.0cm]{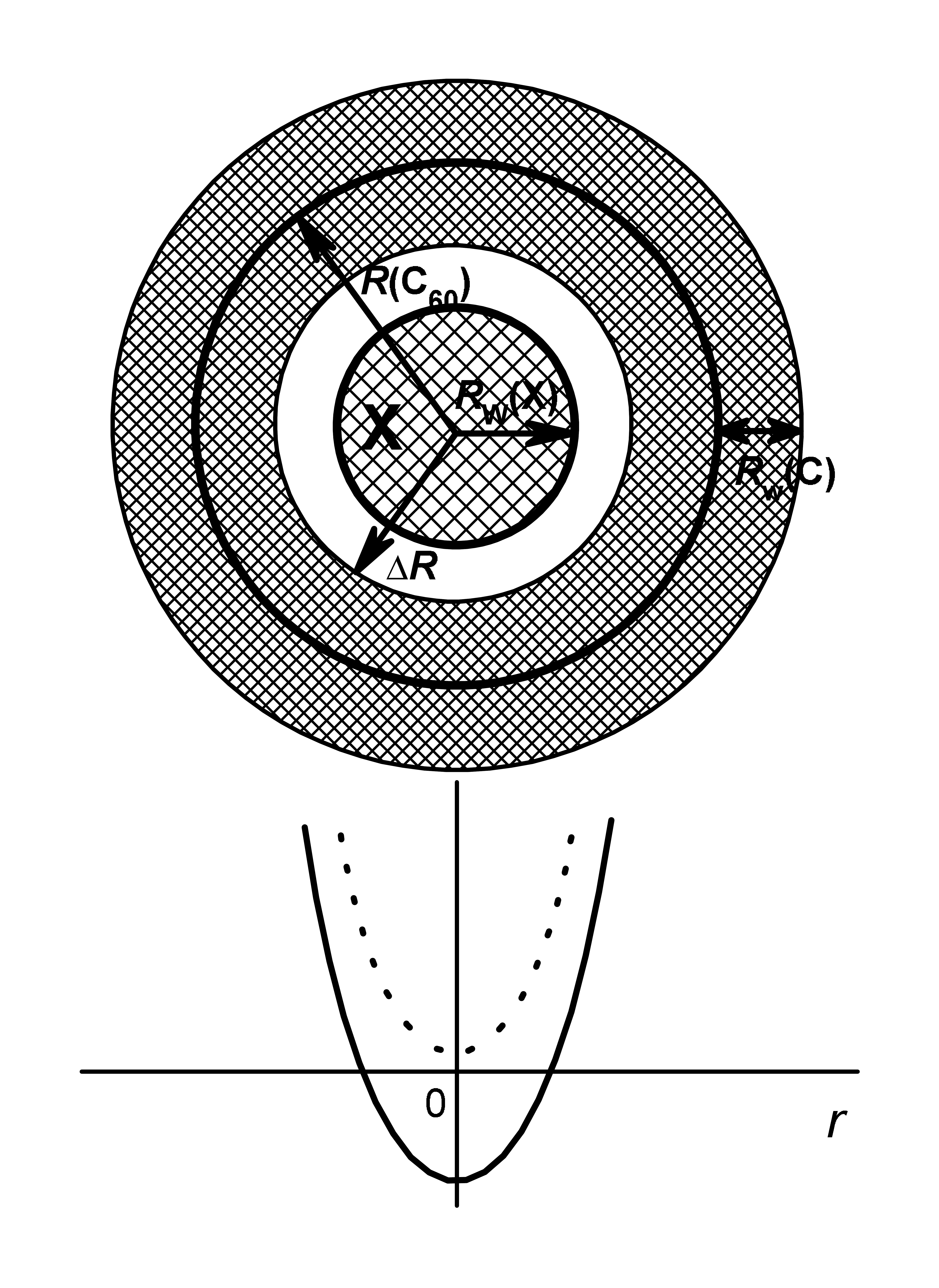}
\caption{Cross-section of an endohendral fullerene
$\texttt{X@C}_{60}$. We consider the carbon cage as a spherical
layer of thickness $2R_{\texttt{W}}(\texttt{C})$ and central
radius $R(\texttt{C}_{60})$. An atom $\texttt{X}$ is a central
nobel gas atom of van der Waals radius
$R_{\texttt{W}}(\texttt{X})$ placed into the inner cavity of
radius $\Delta R$. In lower part of the figure the potential
energy of the noble gas atom in $\texttt{C}_{60}$ is shown
schematically as a function of displacement from the center. For
$\texttt{He}$, $\texttt{Ne}$, $\texttt{Ar}$ the energy are shown
by a solid line. We can see an energy gain. For $\texttt{Kr}$,
$\texttt{Xe}$ the energy are shown by a dotted line. We can see
positive energy of the atoms in the fullerene.} \label{Fig1}
\end{figure}

Noble gas atom interacts with carbon cage by van der Waals
interaction. As it was explained in \cite{elets,weid,piet} on an
example of a complex $\texttt{N@C}_{60}$ (nitrogen like a noble
gas atom does not make a covalent bond with a carbon cage, however
it has nonzero spin) a fullerene has inner cavity in its center -
Fig.\ref{Fig1}. Size of the inner cavity is $\Delta
R=R(\texttt{C}_{60})-R_{\texttt{W}}(\texttt{C})=1.87\texttt{A}$,
where $R(\texttt{C}_{60})=3.57\texttt{A}$ is a radius of
fullerene, $R_{\texttt{W}}(\texttt{C})=1.70\texttt{A}$ is van der
Waals radius of a carbon atom. An atom $\texttt{X}$ can be placed
into the inner cavity if $R_{\texttt{W}}(\texttt{X})\lesssim\Delta
R$. Van der Waals radii of noble gas atoms are
$R_{\texttt{W}}(\texttt{He})=1.40\texttt{A}$,
$R_{\texttt{W}}(\texttt{Ne})=1.54\texttt{A}$,
$R_{\texttt{W}}(\texttt{Ar})=1.88\texttt{A}$,
$R_{\texttt{W}}(\texttt{Kr})=2.02\texttt{A}$,
$R_{\texttt{W}}(\texttt{Xe})=2.16\texttt{A}$. Thus $\texttt{He}$,
$\texttt{Ne}$, $\texttt{Ar}$ can be placed into fullerene
\cite{breton}. If $R_{\texttt{W}}(\texttt{X})\lesssim\Delta R$
then van der Waals attraction acts between noble gas atom and the
carbon cage. Therefore when we place atoms $\texttt{He}$,
$\texttt{Ne}$, $\texttt{Ar}$ in the cage we have an energy gain,
and when we place atoms $\texttt{Kr}$, $\texttt{Xe}$ in the cage
we increase energy of the system \cite{jime}.

The simplest endohendral fullerene is $\texttt{He@C}_{60}$. In a
work \cite{patch} a potential energy of $\texttt{He@C}_{60}$
relative to noninteracting $\texttt{He}$ and $\texttt{C}_{60}$ as
a function of the $\texttt{He}$ displacement from the center of
$\texttt{C}60$ has been calculated with density functional method
assuming no relaxation of the cage atoms. For two model potentials
they obtained:
\begin{eqnarray}
  V_{1}(r) &=& (0.8098r^{6}+0.8428r^{4}+1.905r^{2}-2.001)\texttt{kcal/mol} \label{1.1}\\
  V_{2}(r) &=&
  (0.6706r^{6}+0.5367r^{4}+1.370r^{2}-1.999)\texttt{kcal/mol},
  \label{1.2},
\end{eqnarray}
where $r$ is a displacement of a helium atom from the center of
$\texttt{C}_{60}$ (in angstroms). Thus the energy gain for
$\texttt{He}$ inside $\texttt{C}_{60}$ is $1000\texttt{K}$ (we
will use a system of units where $\hbar=k_{\texttt{B}}=1$). If the
potential (\ref{1.1},\ref{1.2}) are consider as harmonic, then
corresponding frequencies and corresponding oscillator lengthes
are
\begin{eqnarray}
    &&\omega_{1}=151\texttt{K}, \quad \xi_{1}=0.28\texttt{A}\label{1.3}\\
    &&\omega_{2}=128\texttt{K}, \quad \xi_{2}=0.31\texttt{A}.\label{1.4}
\end{eqnarray}
The oscillator lengthes are average displacements of a helium atom
from the center of $\texttt{C}_{60}$. Since $\xi_{1,2}^{2}\gg
\xi_{1,2}^{4}\gg \xi_{1,2}^{6}$ then we can keep the quadratic
terms in Eqs.(\ref{1.1},\ref{1.2}) only. The potential energy
energy can be measured from the bottom of the potential well:
$V_{1,2}(0)=0$. This means harmonic approximation for potential
energy of a helium atom in a center of fullerene. Then energy of
the atom in a center of a fullerene is energy of a 3D harmonic
oscillator
\begin{equation}\label{1.5}
    E=\omega\left(2n+l+\frac{3}{2}\right),
\end{equation}
where $n$ is a main quantum number, $l$ is an orbital quantum
number, and besides a magnetic quantum number is $m=-l,-l+1,\ldots
l-1,l$. States of the noble gas atom in a center of a fullerene
are states of a 3D harmonic oscillator:
$\Psi_{n,l,m}(r,\theta,\varphi)=R_{n,l}(r)Y_{l,m}(\theta,\varphi)$,
where $R_{n,l}(r)$ is a radial function, $Y_{l,m}(\theta,\varphi)$
is an orbital wave function.

\subsection {Interaction with electrons.}\label{impur2}

\begin{figure}[h]
\includegraphics[width=7.0cm]{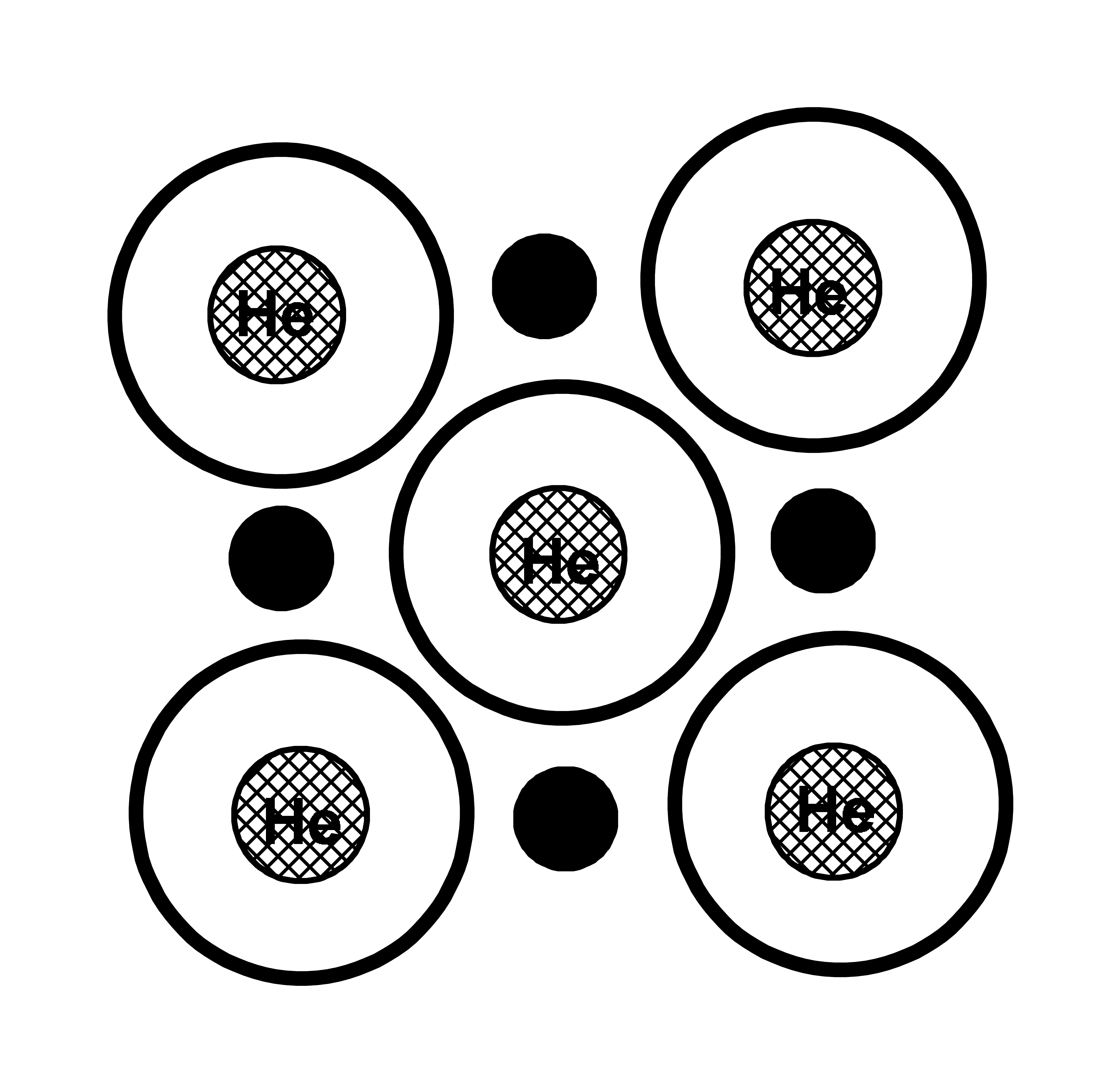}
\caption{Plane (100) face-centered cubic lattice of
$\texttt{K}_{3}\texttt{He@C}_{60}$. Big circles are fullerene
molecules, small black circles are alkali ions, circles in center
of the fullerene are helium atoms.} \label{Fig2}
\end{figure}

Let the endohendral fullerene $\texttt{He@C}_{60}$ is embeded into
into electron gas in a conductor. The most suitable system for our
aim is an alkali-doped fulleride, for example
$\texttt{K}_{3}\texttt{C}_{60}$. In alkali-doped fullerides the
molecules $\texttt{C}_{60}$ are not impurities but they is placed
regularly in a crystal lattice. A lattice constant is
$a=14.23\texttt{A}$. There are 4 molecules of fullerene per a
lattice cell. In each fullerene we can placed helium atom, then we
have a substance $\texttt{K}_{3}\texttt{He@C}_{60}$ -
Fig.(\ref{Fig2}), where the helium atoms can be considered as
impurities (justification is in Section \ref{super}). Electrons
interacts with a carbon cage and with a central helium atom of the
endohendral fullerene. The carbon cage is made with $\sigma$-bonds
of $1\texttt{s}$ electrons of carbon atoms (very small overlap),
$\sigma$-bonds of $2\texttt{sp}^{2}$ hybridized electrons and
$\pi$-bonds of 60 $2\texttt{p}$ electrons distributed over 8
Huckel orbits due truncated icosahedron symmetry of a fullerene.
In the molecule there are the highest occupied molecular orbital
(HOMO) $\texttt{h}_{\texttt{u}}$ occupied by 10 electrons with
energy $\approx -5.8\texttt{eV}$ and the lowest unoccupied
molecular orbital (LUMO) $\texttt{t}_{1\texttt{u}}$ which can
carry 6 electrons with energy $\approx -4.2\texttt{eV}$
\cite{koch}. An overlap of empty $\texttt{t}_{1\texttt{u}}$ levels
of neighbour fullerenes forms a conduction band of width $\approx
0.5\texttt{eV}$ \cite{lieber,forro,haddon}. Each alkali metal atom
give one electron so that each fullerene obtains three electron,
that is $\texttt{t}_{1\texttt{u}}$ level is filled to half.
Accordingly the conduction band is filled to half too and the
alkali doped fulleride is a conductor \cite{gunn}.

Let us consider intramolecular interaction of an excess electron
on $\texttt{t}_{1\texttt{u}}$ level (a conduction electron) with
oscillations of the endohendral fullerene $\texttt{He@C}_{60}$. To
describe the interaction we use \textit{orthogonalized plane wave
method} and \textit{pseudopotential method} \cite{anim}. This
method is applied for description of a conduction electron in a
metal, where state of the electron is a plane wave
$\frac{1}{V}e^{i\textbf{kr}}$ between ions and \textit{rapidly
oscillating function} $\psi_{n,l,m}(\textbf{r})$ (wave function of
a valent atomic orbital) near each ion. \textit{The oscillations
increase kinetic energy of an electron near an ion. The increasing
of kinetic energy acts in the vicinity of the ion cores like some
repulsive potential}. Huckel orbits of an endohendral fullerene
$\texttt{He@C}_{60}$ is formed from overlap of orbits of a carbon
cage ($\sigma$-bonds and $\pi$-bonds) and $1s$-state of a helium
atom. As a result we have new orbits, but, in consequence of the
small overlap of orbits of $\texttt{C}_{60}$ and $\texttt{He}$,
energy of electron's states is changed little. However fact of the
overlap influences on states of conduction electrons as follows.
The conduction band is formed of overlap of the valent
$\texttt{t}_{1\texttt{u}}$ states of neighboring molecules. As for
atoms in the orthogonalized plane wave method, the wave function
$\varphi$ of a conduction electron with energy $E$ are orthogonal
to the occupied states, in particular to $1\texttt{s}$-state of an
electron in the helium atom $\chi$:
\begin{eqnarray}
    &&\langle\varphi|\chi\rangle=0,\quad \chi=\frac{1}{\sqrt{\pi
    a^{3}}}e^{-r/a}\label{2.1}
\end{eqnarray}
where $a=0.31\texttt{A}$ is atomic radius,
$E_{\chi}=-38.83\texttt{eV}$ is energy of the electron in
$1\texttt{s}$-state. The orthogonality takes into consideration
above-mentioned rapid oscillations of the conduction electron's
wave function near the molecular core. The rapid oscillations give
an additional kinetic energy
$\sim\int_{0}^{a}\varphi^{+}\Delta\varphi d\textbf{r}$ in vicinity
of the molecular cores. The pseudopotential method allows us to
rewrite the kinetic energy as some effective potential
$U_{\texttt{eff}}$ - pseudopotential, and unknown exact wave
function of a valent electron can be replaced by some pseudowave
function $\psi$ which has not rapid oscillations near a molecular
(ionic) core, but this state has energy which is the same as in
the exact state - $E$. Then according to \cite{anim} the effective
potential can have a form
\begin{equation}\label{2.6}
   \widehat{U}_{\texttt{eff}}=\widehat{U}+\left(E-E_{\chi}\right)|\chi\rangle\langle\chi|\equiv\widehat{U}+\widehat{U}_{\texttt{ps}}.
\end{equation}
The second part of the potential is repulsive because energy of a
valent electron $E$ is more then energy of ionic core
$E>E_{\chi}$. The repulsive pseudopotential is manifestation of
the Pauli principle prohibiting the valence electrons to be in
region of the occupied orbitals (in our case in region of the
occupied $1\texttt{s}$-orbital of a helium atom). Mathematically
this fact is expressed by the orthogonality (\ref{2.1}).

In coordinate representations the effect on a wave function and an
quantum-mechanical average of the pseudopotential are
correspondingly:
\begin{eqnarray}
    && \widehat{U}_{\texttt{ps}}|\psi\rangle=\left(E-E_{\chi}\right)\chi(\textbf{r})\int\chi(\textbf{r}')\psi(\textbf{r}')d\textbf{r}'\label{2.7}\\
&&\langle\psi|\widehat{U}_{\texttt{ps}}|\psi\rangle=\left(E-E_{\chi}\right)
\int\chi(\textbf{r})\psi(\textbf{r})d\textbf{r}\int\chi(\textbf{r}')\psi(\textbf{r}')d\textbf{r}'\label{2.8}
\end{eqnarray}
We can see the pseudopotential is a nonlocal operator. The
function $\chi(\textbf{r})$ plays a role of a formfactor. Using of
the nonlocal operator is problematical. However the nonlocal
pseudopotential can be localized taking into account the effects
of nonlocality. We propose a following method. Let us write the
localized pseudopotential in a form (we take out the function
$\psi$ under the integral sign in Eq.(\ref{2.7})):
\begin{equation}\label{2.9}
    \widehat{U}_{\texttt{loc}}=A\left(E-E_{\chi}\right)\chi(\textbf{r})\int\chi(\textbf{r}')d\textbf{r}'=8A\left(E-E_{\chi}\right)e^{-\frac{r}{a}},
\end{equation}
where a constant $A$ is found from equality of quantum-mechanical
averages:
\begin{eqnarray}
    \langle\psi|\widehat{U}_{\texttt{ps}}|\psi\rangle=\langle\psi|\widehat{U}_{\texttt{loc}}|\psi\rangle.\label{2.10}
\end{eqnarray}
Thus constant $A$ considers nonlocality of the pseudopotential. In
the first approximation the pseudowave function can be chosen as a
plane wave
$\psi=\frac{1}{\sqrt{V}}e^{i\textbf{kr}}\equiv|\textbf{k}\rangle$.
Then we have
\begin{eqnarray}
    &&\langle\textbf{k}|\widehat{U}_{\texttt{ps}}|\textbf{k}\rangle=\frac{64\pi}{V}\left(E-E_{\chi}\right)a^{3}A\label{2.11}\\
    &&\langle\textbf{k}|\widehat{U}_{\texttt{loc}}|\textbf{k}\rangle=\frac{64\pi}{V}\left(E-E_{\chi}\right)a^{3}\left(1+a^{2}k^{2}\right)^{-4}.\label{2.12}
\end{eqnarray}
Hence $A=\left(1+a^{2}k^{2}\right)^{-4}$, where we can suppose
$k=k_{\texttt{F}}$. As one would expect the nonlocality somewhat
weakens the interaction. The localized pseudopotential in
coordinate space and in momentum spaces is
\begin{eqnarray}
    &&\widehat{U}_{\texttt{loc}}(r)=\frac{8\left(E-E_{\chi}\right)}{\left(1+a^{2}k_{\texttt{F}}^{2}\right)^{4}}e^{-\frac{r}{a}}\label{2.13}\\
    &&\widehat{U}_{\texttt{loc}}(q)=\int e^{-i\textbf{kr}}\widehat{U}_{\texttt{loc}}(r)d\textbf{r}
    =\frac{64\pi\left(E-E_{\chi}\right)}{\left(1+a^{2}k_{\texttt{F}}^{2}\right)^{4}}\frac{a^{3}}{\left(1+a^{2}q^{2}\right)^{2}}.\label{2.14}
\end{eqnarray}

Let us consider the potential $U$ in Eq.(\ref{2.6}). In the
pseudopotential method for a metal the potential is Coulomb
$-\frac{e^{2}Z}{r}$, where $Z\neq 0$ is a degree of ionization of
an atom. Then contribution of the first term in Eq.(\ref{2.6}) is
negative, contribution of the pseudopotential is positive. Thus
compensation of these two terms takes place. This leads to a weak
resulting potential $U_{\texttt{eff}}$ \cite{anim}. In our case a
helium atom in an endohedral fullerene is neutral $Z=0$. Thus the
compensation does not take place and the pseudopotential is a
strong repulsive potential unlike a metal. Interaction of an
electron with an atom is
\begin{equation}\label{2.15}
    U(q)=-\frac{4\pi
    e^{2}}{q^{2}}(Q-n_{\textbf{q}})=-\frac{2e^{2}}{q^{2}}\left(1-\left(1+\frac{a^{2}}{8}q^{2}\right)^{-2}\right),
\end{equation}
where $Q$ is a nuclear charge and $n_{\textbf{q}}$ is Fourier
transform of electron density \cite{vakar}. The potential for a
neutral helium atom is weak short range. Estimation of values of
the pseudopotential (\ref{2.14}) and the potential (\ref{2.15})
shows $U_{\texttt{loc}}\gg U$, so that the potential $U$ can be
omitted and $U_{\texttt{eff}}=U_{\texttt{loc}}$.

\section {Superconductivity}\label{super}
\subsection {General equations}\label{super1}

Conduction electrons interact with intramolecular
$\texttt{H}_{\texttt{g}}(1)-\texttt{H}_{\texttt{g}}(8)$ phonons
which have frequencies  $273-1575\texttt{cm}^{-1}$ \cite{gunn}.
The interaction results to superconductivity of alkali-doped
fullerenes. So, in a substance $\texttt{K}_{3}\texttt{C}_{60}$ a
critical temperature is $T_{\texttt{C}}=19.3\texttt{K}$. In this
section we consider an influence of interaction with a helium atom
in an endohendral fullerene $\texttt{He}@\texttt{C}_{60}$ on
superconducting properties of alkali doped fulleride. In other
words our aim is to find the critical temperature of a substance
$\texttt{K}_{3}\texttt{He}@\texttt{C}_{60}$.

Let an electron moves in a field created by $N$ scatterers
(impurities) which are placed in points $\textbf{R}_{j}$ by a
random manner with concentration $\rho=\frac{N}{V}$. Such system
is spatially inhomogeneous, hence momentum of a quasiparticle  is
not conserved. However we can use mean-field approximation, that
is effect of all impurities is replaced by a mean field using an
averaging operation over a disorder. The operation for a
propagator of a quasiparticle has a form \citep{levit}:
\begin{eqnarray}\label{3.1}
\left\langle
G(x,x')\right\rangle=-i\left\langle\frac{\left\langle\widehat{T}\psi^{+}(x)\psi(x')\widehat{U}\right\rangle_{0}}
{\left\langle\widehat{U}\right\rangle_{0}}\right\rangle_{\texttt{disorder}},
\end{eqnarray}
where $\widehat{U}$ is an evolution operator,
$\langle\ldots\rangle_{0}$ is done over a ground state of Fermi
system and a lattice (in the numerator and the denominator
separately). The averaging over the disorder is done in the
following way - at first the propagator is calculated at the given
disorder, and only then the averaging $\langle\ldots\rangle$ is
done (the whole fraction is averaged). The averaging over an
ensemble of samples with all possible positions of impurities
recovers spatial homogeneity of the system, hence quasiparticles'
momentums are conserved. Practically the averaging
$\left\langle\right\rangle_{\texttt{disorder}}$ is done as
follows:
\begin{equation}\label{3.2}
\sum_{j}\textbf{R}_{j} \longrightarrow\rho\int d\textbf{R},
\end{equation}
where $\rho$ is concentration of impurities. Thus the impurities
are "smeared" over the system with concentration $\rho$ and they
act on quasiparticles as a mean field. Diagram technique for
disordered system if interaction of quasiparticles with impurities
is retarded has been developed in a work \cite{grig2}.

On the other hand we propose a substance
$\texttt{K}_{3}\texttt{He}@\texttt{C}_{60}$, where the molecules
$\texttt{He}@\texttt{C}_{60}$ is not impurities but they is placed
regularly in a crystal lattice - Fig(\ref{Fig2}). This system can
be considered in an approximation of a jelly model. In this model
the helium atoms are "smeared" over the system like electron and
ion subsystems in a metal and they act on quasiparticles as a mean
field. Thus the jelly model leads to the same result as the
averaging (\ref{3.1}) in the disordered metal with concentration
of impurities $\rho=4/a^{3}$, where $a=14.23\texttt{A}$ is a
lattice constant and 4 molecules of fullerene per a lattice cell
are. However there is a significant difference of these model. As
it is well known impurities in a metal result to decreasing of
density of states at Fermi surface up to Anderson localization if
density of the impurities is large \cite{sad,sad1}. The decreasing
is result of scattering of conduction electrons by randomly
distributed impurities. In
$\texttt{K}_{3}\texttt{He}@\texttt{C}_{60}$ we have a regular
arrangement of helium atoms that does not result to decreasing of
the density of states on Fermi surface and Anderson localization.

Let us calculate critical temperature $T_{\texttt{C}}^{\ast}$ of a
system $\texttt{K}_{3}\texttt{He}@\texttt{C}_{60}$, where helium
atoms play a role of an impurity. Critical temperature of a pure
system $\texttt{K}_{3}\texttt{C}_{60}$ is
$T_{\texttt{C}}=19.3\texttt{K}$. In a work \cite{grig2} Eliashberg
equations has been generalized for a system metal+impurities. The
impurity is a harmonic oscillator with transition frequencies
between any of its states $\phi_{B}$ and $\phi_{A}$:
$\omega_{AB}=E(B)-E(A)$, interaction of a conduction electron with
the impurity is $U(r)$, the impurities are distributed over the
system with concentration $\rho$. The conduction electrons
interact with phonon with frequency $\omega_{\texttt{D}}$,
electron-phonon coupling constant is $g$. The equations have a
form
\begin{eqnarray}
&&Z(\varepsilon_{n})\Delta_{n}=
\sum_{m=-\infty}^{+\infty}\left(L(n-m)-\mu^{\ast}\right)
\frac{\pi T\widetilde{\Delta}_{m}}{\sqrt{\widetilde{\varepsilon}_{m}^{2}+|\widetilde{\Delta}_{m}}|^{2}}\label{3.3}\\
&&(1-Z(\varepsilon_{n}))\varepsilon_{n}=\sum_{m=-\infty}^{+\infty}L(n-m)\frac{\pi
T\widetilde{\varepsilon}_{m}}{\sqrt{\widetilde{\varepsilon}_{m}^{2}+|\widetilde{\Delta}_{m}}|^{2}}\label{3.4},\\
&&\widetilde{\Delta}_{n}=\Delta_{n}+
\sum_{m=-\infty}^{+\infty}W(n-m)\frac{\pi T\widetilde{\Delta}_{m}}{\sqrt{\widetilde{\varepsilon}_{m}^{2}+|\widetilde{\Delta}_{m}}|^{2}}\label{3.5}\\
&&\widetilde{\varepsilon}_{n}=\varepsilon_{n}+\sum_{m=-\infty}^{+\infty}W(n-m)\frac{\pi
T\widetilde{\varepsilon}_{m}}{\sqrt{\widetilde{\varepsilon}_{m}^{2}+|\widetilde{\Delta}_{m}}|^{2}}\label{3.6},
\end{eqnarray}
where $\varepsilon_{n}=\pi T(2n+1)$ is an energetic parameter in
Matzubara representation, $\Delta_{n}$ is a superconducting gap,
$\mu^{\ast}$ is Coulomb pseudopotential, $\widetilde{\Delta}$ is a
renormalized gap and $\widetilde{\varepsilon}_{n}$ is a
renormalized energy parameter. The renormalization takes place due
scattering of conduction electrons by impurities. Electron-phonon
and electron-impurity coupling is represented by functions
\begin{eqnarray}
&&L(n-m)=g\frac{\omega_{\texttt{D}}^{2}}{(n-m)^{2}\pi^{2}T^{2}+\omega_{\texttt{D}}^{2}}\label{3.7}\\
&&W(n-m)=
\sum_{A}\sum_{B}\varpi_{A}\int_{0}^{2k_{F}}\int_{0}^{\pi}\frac{2\rho
qdq\sin\theta
d\theta}{\omega_{AB}v_{F}(2\pi)^{2}}\left|U(\textbf{q})\langle
B|A\rangle_{\textbf{q}}\right|^{2}\frac{\omega_{AB}^{2}}{(n-m)^{2}\pi^{2}T^{2}+\omega_{AB}^{2}},\label{3.8}
\end{eqnarray}
where $v_{F}$ and $k_{F}$ are Fermi velocity and Fermi momentum
accordingly,
\begin{eqnarray}
    &&\langle B|A\rangle_{\textbf{q}}=\int
e^{-i\textbf{qr}}\phi_{B}^{+}\phi_{A}d\textbf{r},\label{3.10}
\end{eqnarray}
In a case of nonzero temperature $T\neq 0$ the impurities are
distributed over states $|A\rangle,|B\rangle,|C\rangle,\ldots$
with probability
\begin{equation}\label{3.8a}
    \varpi_{A}=\frac{1}{Z}\exp\left(-\frac{E_{A}-E_{0}}{T}\right),\quad Z=\sum_{A}\exp\left(-\frac{E_{A}-E_{0}}{T}\right)
\end{equation}
where $E_{0}$ is a  ground state energy of an impurity, and the
summation is extended on all possible states (we use a system of
units where $\hbar=k_{\texttt{B}}=1$). If impurities are harmonic
oscillators then we can simplify Eq.(\ref{3.8}):
\begin{eqnarray}
W(n-m)\approx \int_{0}^{2k_{F}}\int_{0}^{\pi}\frac{2\rho
qdq\sin\theta
d\theta}{\omega_{AB}v_{F}(2\pi)^{2}}\left|U(\textbf{q})\langle
B|A\rangle_{\textbf{q}}\right|^{2}\frac{\omega_{AB}^{2}}{(n-m)^{2}\pi^{2}T^{2}+\omega_{AB}^{2}},\label{3.8b}
\end{eqnarray}
where $|A\rangle$ is a ground state of the oscillator
$(n=0,l=0,m=0)$, $|B\rangle$ is the nearest excited state
$(n=0,l=1,m=0)$. Justification of the approximation (\ref{3.8b})
is given in Section \ref{calc} in numerical calculations.

The gap $\widetilde{\Delta}_{m}$ is an even function of $2m+1$,
but the energy parameter $\widetilde{\varepsilon}_{m}$ is an odd
function of $2m+1$. Hence these functions are renormalized in
different ways:
\begin{equation}\label{3.11}
\frac{\widetilde{\Delta}}{\widetilde{\varepsilon}}>\frac{\Delta}{\varepsilon}.
\end{equation}
This unequality ensures increasing of the gap $\Delta$ as compared
with a pure superconductor or with a dirty superconductor with
elastic impurities where an equality
$\frac{\widetilde{\Delta}}{\widetilde{\varepsilon}}=\frac{\Delta}{\varepsilon}$
takes place. Thus Anderson theorem is violated in the sense that
embedding of the impurities in $s$-wave superconductor increases
its critical temperature.

The set of equations (\ref{3.3}-\ref{3.6}) can be simplified using
an approximation for an electron-electron interaction amplitude
$gw(\varepsilon_{n},\varepsilon_{n'})$ with a method proposed in
\citep{levit}:
\begin{equation}\label{3.12}
w(\varepsilon_{n},\varepsilon_{m})\equiv
\frac{\omega^{2}}{(\varepsilon_{n}-\varepsilon_{m})^{2}+\omega^{2}}
\longrightarrow
w(\varepsilon_{n})w(\varepsilon_{m})=\frac{\omega}{\sqrt{\varepsilon_{n}^{2}+\omega^{2}}}
\frac{\omega}{\sqrt{\varepsilon_{m}^{2}+\omega}}.
\end{equation}
Here $\omega=\omega_{\texttt{D}}, \omega_{AB},\ldots$ is
characteristic frequency of the interaction. In addition we
suppose the gap to be real $\Delta=\Delta^{+}$ and to depend on
energy as follows:
\begin{equation}\label{3.12a}
    \Delta_{n}=\Delta\frac{\omega_{D}}{\sqrt{\varepsilon_{n}^{2}+\omega_{D}^{2}}}
    \equiv\Delta w_{\texttt{D}}(\varepsilon_{n}).
\end{equation}
Moreover we can consider some effective electron-phonon coupling
constant $g$ instead of $g-\mu^{\ast}$ so that would have the
correct critical temperature $T_{\texttt{C}}=19.3\texttt{K}$ of a
pure $\texttt{K}_{3}\texttt{C}_{60}$ using the approximation
(\ref{3.12}). The approximation (\ref{3.12}) removes contribution
of terms with $m=n$ that is $L(0)$ and $W(0)$. The terms do not
influence upon a gap and a critical temperature and they describe
a scattering of electrons by thermal oscillations of the lattice
and impurities \cite{ginzb}. The thermal oscillations behave like
static impurities with effective concentration
$\rho\frac{2T}{\omega_{0}}$. The scattering gives an additional
contribution in resistance of the metal analogously to a
contribution of thermal phonons. Thus the terms with $m \neq n$
can violate Anderson's theorem only.

Then Eqs.(\ref{3.3}-\ref{3.6}) in the approximation (\ref{3.12})
have a form
\begin{eqnarray}
&&Z(\varepsilon_{n})\Delta_{n}= g\sum_{m=-\infty}^{+\infty}
\frac{\pi
T\widetilde{\Delta}_{m}}{\sqrt{\widetilde{\varepsilon}_{m}^{2}+|\widetilde{\Delta}_{m}}|^{2}}
w_{\texttt{D}}(\varepsilon_{n})w_{\texttt{D}}(\varepsilon_{m})\label{3.13}\\
&&(1-Z(\varepsilon_{n}))\varepsilon_{n}=g\sum_{m=-\infty}^{+\infty}\frac{\pi
T\widetilde{\varepsilon}_{m}}{\sqrt{\widetilde{\varepsilon}_{m}^{2}+|\widetilde{\Delta}_{m}}|^{2}}
w_{\texttt{D}}(\varepsilon_{n})w_{\texttt{D}}(\varepsilon_{m})=0\Longrightarrow Z=1\label{3.14},\\
&&\widetilde{\Delta}_{n}=\Delta_{n}+
G\sum_{m=-\infty}^{+\infty}\frac{\pi
T\widetilde{\Delta}_{m}}{\sqrt{\widetilde{\varepsilon}_{m}^{2}+|\widetilde{\Delta}_{m}}|^{2}}
w_{AB}(\varepsilon_{n})w_{AB}(\varepsilon_{m})\label{3.15}\\
&&\widetilde{\varepsilon}_{n}=\varepsilon_{n}+G\sum_{m=-\infty}^{+\infty}\frac{\pi
T\widetilde{\varepsilon}_{m}}{\sqrt{\widetilde{\varepsilon}_{m}^{2}+|\widetilde{\Delta}_{m}}|^{2}}
w_{AB}(\varepsilon_{n})w_{AB}(\varepsilon_{m})=\varepsilon_{n}+0\label{3.16},
\end{eqnarray}
where $G$ is electron-impurity coupling constant:
\begin{equation}\label{3.17}
    G=\int_{0}^{2k_{F}}\int_{0}^{\pi}\frac{2\rho
qdq\sin\theta
d\theta}{\omega_{AB}v_{F}(2\pi)^{2}}\left|U(\textbf{q})\langle
B|A\rangle_{\textbf{q}}\right|^{2}
\end{equation}
Then Eqs.(\ref{3.13}-\ref{3.16}) can be reduced to a form
\begin{eqnarray}
&&\Delta= g\sum_{m=-\infty}^{+\infty} \frac{\pi
T\widetilde{\Delta}_{m}}{\sqrt{\widetilde{\varepsilon}_{m}^{2}+|\widetilde{\Delta}_{m}}|^{2}}
w_{\texttt{D}}(\varepsilon_{m})\label{3.18}\\
&&\widetilde{\Delta}_{m}=\Delta
w_{\texttt{D}}(\varepsilon_{m})+\Delta
w_{AB}(\varepsilon_{m})\frac{f}{1-h},\label{3.19}
\end{eqnarray}
where
\begin{eqnarray}
&&f=G \sum_{n=-\infty}^{+\infty} \frac{1}{\sqrt{\pi^{2}
T^{2}(2n+1)^{2}+|\widetilde{\Delta}_{n}|^{2}}}w_{\texttt{D}}(\varepsilon_{n})w_{AB}(\varepsilon_{n})\label{3.20}\\
&&h=G \sum_{n=-\infty}^{+\infty} \frac{1}{\sqrt{\pi^{2}
T^{2}(2n+1)^{2}+|\widetilde{\Delta}_{n}|^{2}}}w_{AB}^{2}(\varepsilon_{n}).\label{3.21}
\end{eqnarray}

Our aim is to find critical temperature of the system. Then we
have to suppose
$\widetilde{\Delta}(T_{\texttt{C}})=\Delta(T_{\texttt{C}})=0$ in
Eqs.(\ref{3.18},\ref{3.19}). In this case Eq.(\ref{3.18}) has form
\begin{eqnarray}\label{3.22}
\Rightarrow 1=g^{2}\sum_{n=-\infty}^{+\infty}
\frac{1}{|2n+1|}\left[w_{\omega_{\texttt{D}}}^{2}(\varepsilon_{n})+
w_{\omega_{\texttt{D}}}(\varepsilon_{n})w_{\omega_{AB}}(\varepsilon_{n})
\frac{G\Upsilon\left(\frac{\omega_{\texttt{D}}}{\pi
T},\frac{\omega_{AB}}{\pi T}\right)}
{1-G\Xi\left(\frac{\omega_{AB}}{\pi T}\right)} \right],
\end{eqnarray}
where
\begin{eqnarray}
\Upsilon\left(\frac{\omega_{\texttt{D}}}{\pi
T},\frac{\omega_{AB}}{\pi T}\right)&=&
\sum_{n=-\infty}^{+\infty}\frac{1}{|2n+1|}\frac{\omega_{\texttt{D}}/\pi
T}{\sqrt{|2n+1|^{2}+(\omega_{\texttt{D}}/\pi T)^{2}}}
\frac{\omega_{AB}/\pi T}{\sqrt{|2n+1|^{2}+(\omega_{AB}/\pi T)^{2}}}\label{3.23}\\
\Xi\left(\frac{\omega_{AB}}{\pi T}\right)&=&\frac{(\omega_{AB}/\pi
T)^{2}}{|2n+1|^{2}+(\omega_{AB}/\pi T)^{2}}\nonumber\\
&=&\left[\gamma+2\texttt{ln}2+
\frac{1}{2}\Psi\left(\frac{1}{2}-\frac{i}{2}\frac{\omega_{AB}}{\pi
T}\right)+\frac{1}{2}\Psi\left(\frac{1}{2}+\frac{i}{2}\frac{\omega_{AB}}{\pi
T}\right)\right].\label{3.24}
\end{eqnarray}
Here $\Psi$ is a digamma function , $\gamma\approx 0.577$ is Euler
constant. Transition temperature $T_{\texttt{C}}^{\ast}$ of the
system metal+impurity is temperature when equality (\ref{3.22}) is
satisfied. If the impurities are absent $G=0$ we have an equation
for critical temperature $T_{\texttt{C}}$ of a pure metal:
\begin{eqnarray}\label{3.25}
\Rightarrow 1=g^{2}\sum_{n=-\infty}^{+\infty}
\frac{1}{|2n+1|}w_{\omega_{\texttt{D}}}^{2}(\varepsilon_{n}),
\end{eqnarray}
We can see that the right side of Eq.(\ref{3.22}) has a
singularity when
\begin{eqnarray}\label{3.26}
1=G\Xi\left(\frac{\omega_{AB}}{\pi T}\right),
\end{eqnarray}
Temperature, when equality (\ref{3.26}) is satisfied, is named
\textit{singularity temperature} $T^{\ast}$ introduced in
\cite{grig1}. In terms of Eqs.(\ref{3.3}-\ref{3.6}) the
singularity temperature is determined by homogeneous set of
equations obtained from Eq.(\ref{3.8}) omitting $\Delta_{n}$:
\begin{eqnarray}\label{3.27}
\sum_{m}W(n-m)\frac{\widetilde{\Delta}_{m}}{|\varepsilon_{m}|}-\widetilde{\Delta}_{n}=0
\end{eqnarray}
A determinant of the set of equations (\ref{3.27}) must be equal
to zero:
\begin{eqnarray}\label{3.28}
\texttt{det}D_{mn}(T_{\texttt{C}}^{\ast})=0,\quad
D_{mn}=\frac{W(n-m)}{|\varepsilon_{m}|} -\delta_{mn},
\end{eqnarray}
where $\delta_{mn}=1$ if $m=n$, $\delta_{mn}=0$ if $m\neq n$. If
an interaction with impurities is nonretarded (elastic):
$W(n-m)=W(0)\delta_{mn}$, then the singularity temperature is
absent \cite{grig2}. Thus Anderson theorem is realized for the
elastic interaction: $T_{\texttt{C}}^{\ast}=T_{\texttt{C}}$. The
singularity temperature is $T^{\ast}<T_{\texttt{C}}^{\ast}$ and it
can be used as a lower estimation of the critical temperature of
the system. Its physical sense is: the singularity temperature is
a superconducting transition temperature if we turn off the
pairing interaction caused by metal's phonon.

Eq.(\ref{3.26}) is simpler than Eq.(\ref{3.28}). Limit cases of
Eq.(\ref{3.26}) are
\begin{eqnarray}\label{3.29}
&&\Xi\left(\frac{\omega_{AB}}{T}\rightarrow
0\right)\rightarrow\frac{7}{4}\zeta(3)\left(\frac{\omega_{AB}}{\pi
T}\right)^{2}\Longrightarrow
T^{\ast}=\frac{\sqrt{7\zeta(3)}}{2\pi}\omega_{AB}\sqrt{G}\nonumber\\
&&\Xi\left(\frac{\omega_{AB}}{T}\rightarrow
\infty\right)\rightarrow\texttt{ln}\left(\frac{2}{\gamma}\frac{\omega_{AB}}{\pi
T}\right)\Longrightarrow
T^{\ast}=\frac{2\omega_{AB}}{\pi\gamma}\exp\left(-\frac{1}{G}\right).
\end{eqnarray}
Thus the limit cases (\ref{3.29}) correspond to limit cases for
critical temperature of a metal superconductor in Eliashberg
equations \cite{mahan}.

\subsection {Calculation of the singularity temperature and the
transition temperature.}\label{calc}

\begin{figure}[h]
\includegraphics[width=12.0cm]{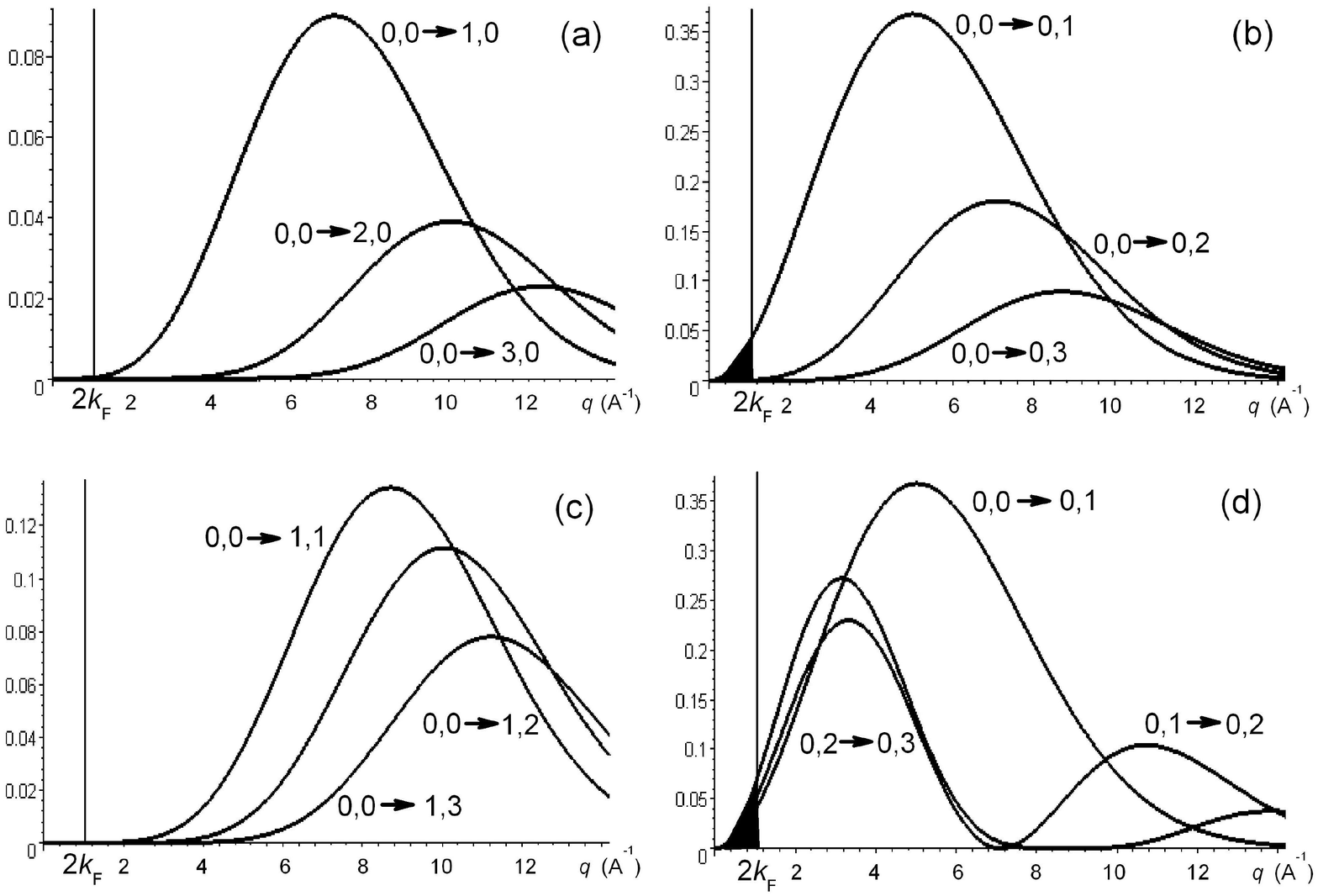}
\caption{Amplitudes $\langle B|A\rangle_{\textbf{q}}$ (\ref{3.10})
for transitions $n,l\rightarrow n',l'$ of 3D harmonic oscillator.
Interval $(0,2k_{\texttt{F}})$ of integration in Eq.(\ref{3.8}) is
painted over.} \label{Fig3}
\end{figure}
\begin{figure}[h]
\includegraphics[width=8.6cm]{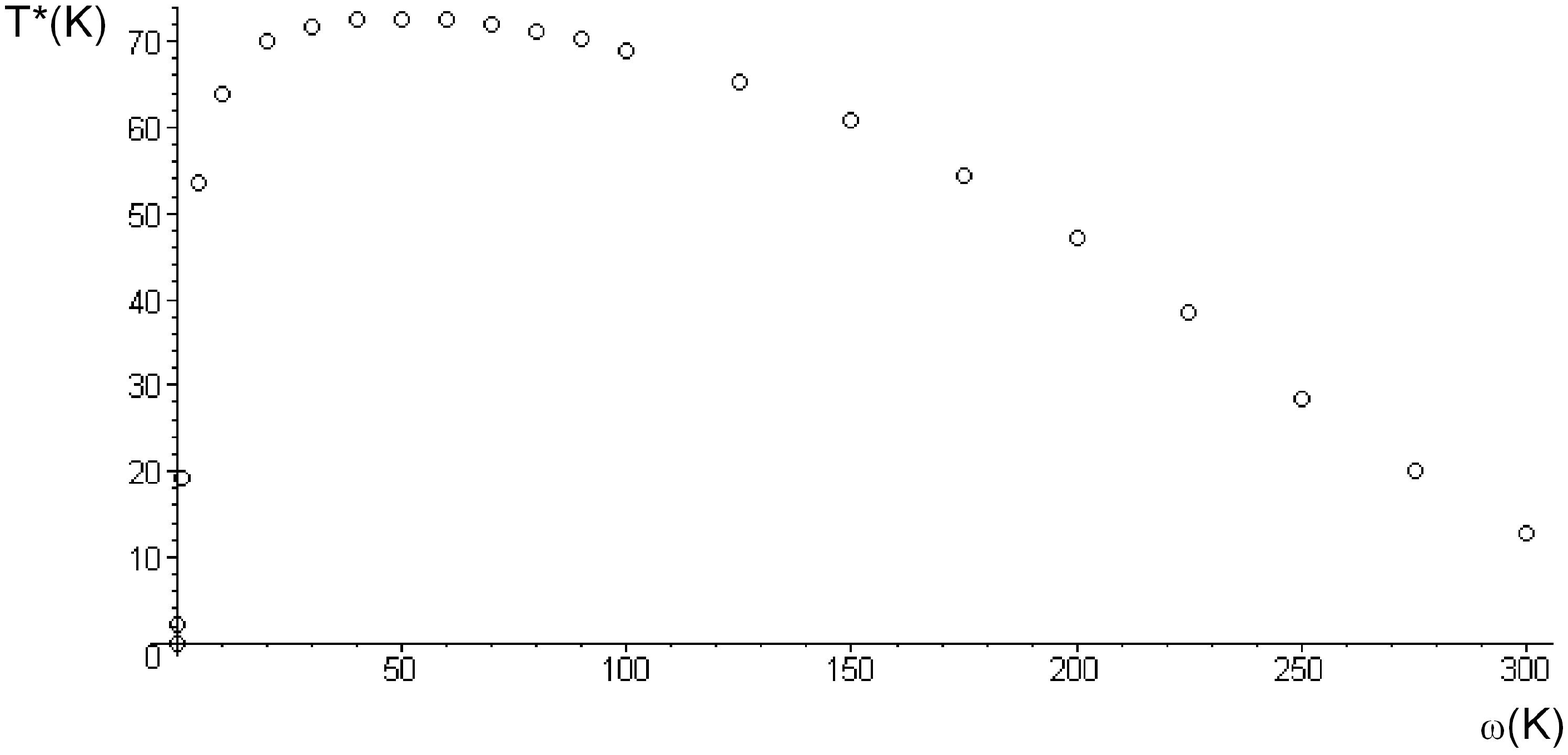}
\caption{Dependence of the singularity temperature $T^{\ast}$ of
$\texttt{K}_{3}\texttt{He}@\texttt{C}_{60}$ on the oscillation
frequency of a helium atom in an endohedral fullerene.}
\label{Fig4}
\end{figure}
\begin{figure}[h]
\includegraphics[width=8.6cm]{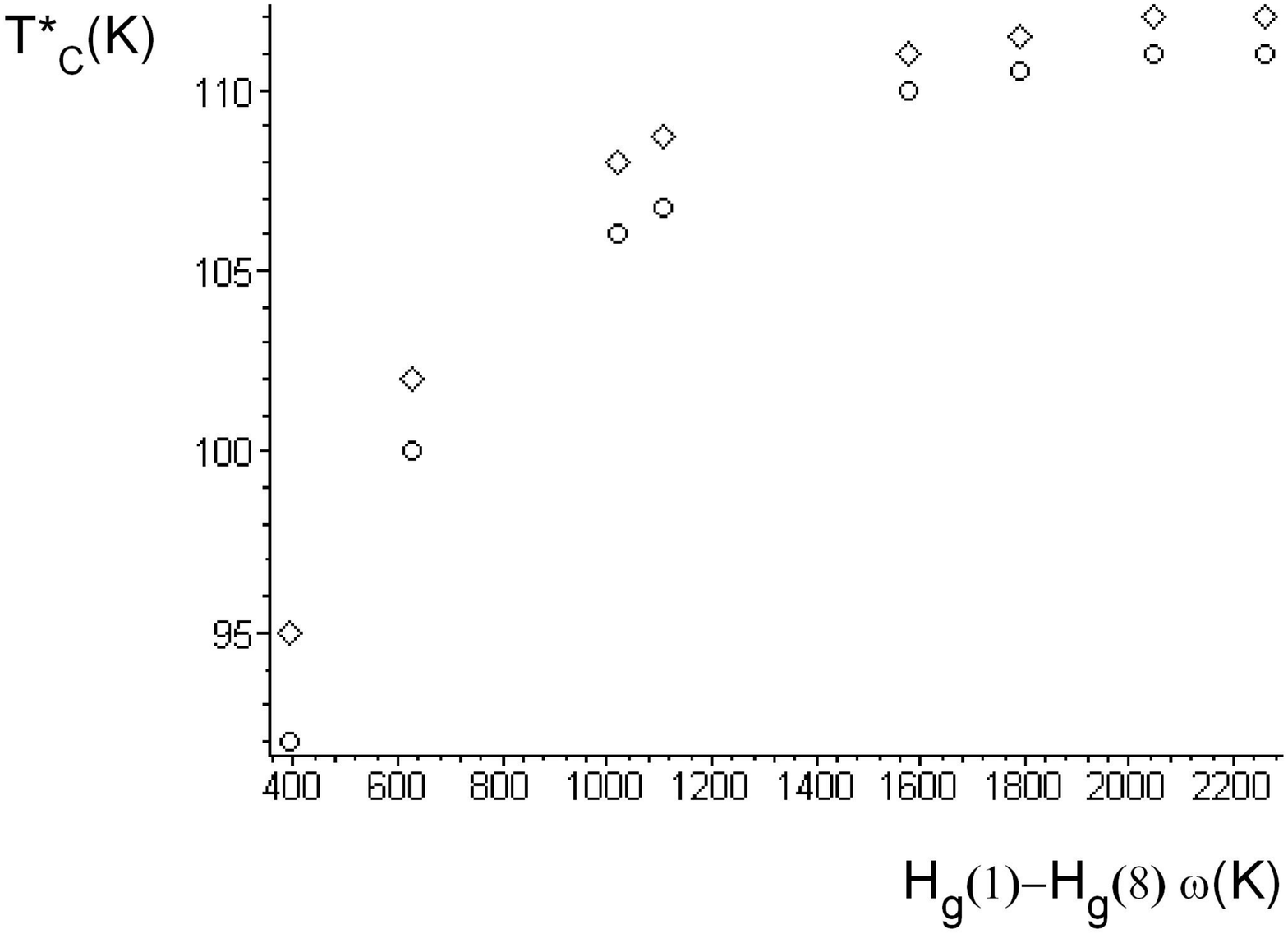}
\caption{The critical temperature $T^{\ast}$ of
$\texttt{K}_{3}\texttt{He}@\texttt{C}_{60}$ for each oscillation
frequency of intramolecular modes
$\texttt{H}_{\texttt{g}}(1)-\texttt{H}_{\texttt{g}}(8)$ of an
fullerene (if only one frequency takes a role in pairing of
electron and all the rest frequencies do not play a part in this
process). Circle symbols - for the frequency of a helium atom in
an endohedral fullerene $\omega_{1}=151\texttt{K}$, diamond
symbols for the frequency $\omega_{1}=128\texttt{K}$.}
\label{Fig5}
\end{figure}

In this section we calculate the singularity temperature and the
transition temperature for
$\texttt{K}_{3}\texttt{He}@\texttt{C}_{60}$ using
Eqs.(\ref{2.14},\ref{3.17},\ref{3.22},\ref{3.26}). Data for
$\texttt{K}_{3}\texttt{C}_{60}$ are given from \cite{varsh}.
First, let us consider the transition amplitude $\langle
B|A\rangle_{\textbf{q}}$ (\ref{3.10}) for various states
$|A\rangle$ and $|B\rangle$ of 3D harmonic oscillator. State of
the oscillator is determined with three quantum numbers $n,l,m$.
Energy of the oscillator is (\ref{1.5}). Ground state is $0,0,0$.
A value (\ref{3.10}) is not equal to zero only for transition when
$\Delta m=0$. Results of the calculation for some transitions are
shown in Fig.\ref{Fig3}. In Eq.(\ref{3.8}) we integrate over
momentum from zero to $2k_{\texttt{F}}=0.92\texttt{A}^{-1}$. This
interval is painted over in Fig.\ref{Fig3}. We can see transitions
with $\Delta n=0, \Delta l=\pm 1$ give contribution in the
integral (\ref{3.8}) only, moreover the contributions of
transitions $l\rightarrow l\pm 1$ are approximately equal for
arbitrary orbital quantum number $l$. For the harmonic oscillator
spectrum (\ref{1.5}) energy of the transitions $l\rightarrow l\pm
1$ are equal $\Delta E(\Delta n=0,\Delta l=\pm 1, \Delta
m=0)=\omega$. Thus the approximation (\ref{3.8b}) can be used
where a state $|A\rangle=\Psi_{0,0,0}(r,\theta,\varphi)$ is a
ground state and a state
$|B\rangle=\Psi_{0,1,0}(r,\theta,\varphi)$ is the first excited
state with $n=0,l=1,m=0$, because
\begin{eqnarray}\label{4.1}
&&\int_{0}^{2k_{F}}\int_{0}^{\pi}\frac{2\rho qdq\sin\theta
d\theta}{\omega v_{F}(2\pi)^{2}}\left|U(\textbf{q})\langle
\Psi_{0,1,0}|\Psi_{0,0,0}\rangle_{\textbf{q}}\right|^{2}\frac{\omega^{2}}{(n-m)^{2}\pi^{2}T^{2}+\omega^{2}}\nonumber\\
&&\approx \int_{0}^{2k_{F}}\int_{0}^{\pi}\frac{2\rho qdq\sin\theta
d\theta}{\omega v_{F}(2\pi)^{2}}\left|U(\textbf{q})\langle
\Psi_{n,l\pm
1,m}|\Psi_{n,l,m}\rangle_{\textbf{q}}\right|^{2}\frac{\omega^{2}}{(n-m)^{2}\pi^{2}T^{2}+\omega^{2}},\nonumber\\
&&\sum_{A}\varpi_{A}=1
\end{eqnarray}

Let us calculate the singularity temperature $T^{\ast}$ using
Eqs.(\ref{3.17},\ref{3.26}). In a potential $U(q)$ (\ref{2.14}) we
can suppose $E=E_{4s}=-4.44\texttt{eV}$ - energy of a valent state
of kalium. Dependence of the temperature $T^{\ast}$ on oscillation
frequency of a helium atom in an endohedral fullerene is shown in
Fig.(\ref{Fig4}). We can see the dependence $T^{\ast}(\omega)$
seems to an effectiveness function in \cite{grig1}. For
frequencies (\ref{1.3},\ref{1.4}) from model \cite{patch} we have
results:
\begin{eqnarray}
    &&\omega_{1}=151\texttt{K}, \quad T^{\ast}=60.4\texttt{K}\label{4.2}\\
    &&\omega_{2}=128\texttt{K}, \quad T^{\ast}=65.1\texttt{K}.\label{4.3}
\end{eqnarray}
Let us calculate the critical temperature $T^{\ast}_{\texttt{C}}$.
Conduction electrons interact with intramolecular
$\texttt{H}_{\texttt{g}}(1)-\texttt{H}_{\texttt{g}}(8)$ phonons,
which have frequencies  $391\div2257\texttt{K}$ \cite{gunn}. Each
vibrational mode is characterized by own coupling constant with
electrons. Interaction with the intramolecular phonons results to
superconductivity of alkali-doped fullerenes. In order to
calculate the critical temperature we propose a following method.
We know that critical temperature of
$\texttt{K}_{3}\texttt{C}_{60}$ is
$T_{\texttt{C}}=19.3\texttt{K}$. We can assume that this critical
temperature can be obtained with each vibrational mode of
fullerene separately - when only one frequency takes a role in
pairing of electron and all the rest frequencies do not play a
part in this process. Corresponding coupling constants $g$ can be
calculated by a formula:
\begin{eqnarray}\label{4.4}
\Rightarrow 1=g\sum_{n=-\infty}^{+\infty}
\frac{1}{|2n+1|}w_{\omega_{\texttt{D}}}^{2}(\varepsilon_{n}),
\end{eqnarray}
then we have
\begin{equation}
\begin{tabular}{|c|c|c|c|c|c|c|c|c|}
  \hline
   & $\texttt{H}_{\texttt{g}}(1)$ & $\texttt{H}_{\texttt{g}}(2)$ & $\texttt{H}_{\texttt{g}}(3)$ & $\texttt{H}_{\texttt{g}}(4)$ & $\texttt{H}_{\texttt{g}}(5)$ & $\texttt{H}_{\texttt{g}}(6)$ & $\texttt{H}_{\texttt{g}}(7)$ & $\texttt{H}_{\texttt{g}}(8)$ \\
  \hline
   & 391\texttt{K} & 626\texttt{K} & 1018\texttt{K} & 1110\texttt{K} & 1575\texttt{K} & 1792\texttt{K} & 2047\texttt{K} & 2257\texttt{K} \\
  \hline
  g& 0.320 & 0.280 & 0.250 & 0.240 & 0.225 & 0.22 & 0.21 & 0.205 \\
  \hline
\end{tabular}
\end{equation}
Then with help of Eqs.(\ref{3.17},\ref{3.22}) we can calculate
$T_{\texttt{C}}^{\ast}$ of
$\texttt{K}_{3}\texttt{He}@\texttt{C}_{60}$. Results for two
frequencies (\ref{4.2},\ref{4.3}) are shown in Fig.(\ref{Fig5}).
We can see that the critical temperature is within the interval
$T_{\texttt{C}}^{\ast}=92\div 112\texttt{K}$, that is it varies
little while the frequencies
$\texttt{H}_{\texttt{g}}(1)-\texttt{H}_{\texttt{g}}(8)$  vary
essentially $391\div 2257\texttt{K}$, thus our method is correct.

\section{Conclusion.}\label{concl}

In this paper we have proposed the model of an impurity with
retarded interaction with quasiparticle. As an impurity we suggest
to use endohedral complexes - a helium atom inside
$\texttt{C}_{60}$ cage: $\texttt{He}@\texttt{C}_{60}$. The atom in
the carbon cage is a oscillator with frequency $\sim
150\texttt{K}$, where we assume that oscillations of the central
atom and carbon cage are independent. Endoendral fullerenes with a
noble atom inside ($\texttt{He}$, $\texttt{Ne}$, $\texttt{Ar}$,
$\texttt{Kr}$, $\texttt{Xe}$) has symmetry and chemical properties
as the hollow fullerences. We propose to use an alkali-doped
fullerides ($\texttt{K}_{3}\texttt{C}_{60}$,
$\texttt{Rb}_{3}\texttt{C}_{60}$ etc.) as a conducting matrices
where these impurity can be applied. These matrices allow us to
create a large concentration of impurities $\rho=4/a$ (where $a$
is a lattice constant) without reduction in the density of states
on Fermi surface. Thus we considered a new hypothetical substance
$\texttt{K}_{3}\texttt{He}@\texttt{C}_{60}$, where the helium atom
can be considered as impurities, because the averaging over
disorder (\ref{3.1}) and the jelly model lead to the same result -
mean-field effect on the conduction electrons. We have shown
potential of interaction of the impurities with electrons is a
pseudopotential (\ref{2.6}) which can be localized to a potential
(\ref{2.14}). Based on results of works \cite{grig1,grig2} we have
calculated singularity temperature which is introduced in these
works and it is important characteristic of a system
matrix+impurity, and we have calculated critical temperature of
the system $\texttt{K}_{3}\texttt{He}@\texttt{C}_{60}$ . The
singularity temperature is within the limits $60.4\div
65.1\texttt{K}$ - Fig.(\ref{Fig4}) and the critical temperature is
within the limits $92\div 112\texttt{K}$ - Fig.(\ref{Fig5}). Thus
effect of the impurities on critical temperature of alkali-doped
fulleride is very strong. However, it should be noted the
pseudopotential is ambiguous \cite{anim}, and it must have fitting
parameters. Hence the obtained results must be considered as
evaluation only.

\end{document}